\documentclass[a4paper,showpacs,twocolumn,amsmath,amssymb,floatfix,prb,preprintnumbers,footinbib]{revtex4}
\usepackage{graphicx}
\usepackage{amsmath,amsfonts}

\begin{document}

\title{Adiabatic pumping through a quantum dot with coulomb interactions:\\ 
	A perturbation expansion in the tunnel coupling}

\author{Janine Splettstoesser}
\affiliation{NEST-CNR-INFM \& Scuola Normale Superiore, Piazza dei
        Cavalieri 7, I-56126 Pisa, Italy}
\affiliation{Institut f\"ur Theoretische Physik III,
	Ruhr-Universit\"at Bochum, D-44780 Bochum, Germany}

\author{Michele Governale}
\affiliation{Institut f\"ur Theoretische Physik III,
	Ruhr-Universit\"at Bochum, D-44780 Bochum, Germany}
\author{J\"urgen K\"onig}
\affiliation{Institut f\"ur Theoretische Physik III,
	Ruhr-Universit\"at Bochum, D-44780 Bochum, Germany}
\author{Rosario Fazio}
\affiliation{NEST-CNR-INFM \& Scuola Normale Superiore, Piazza dei
        Cavalieri 7, I-56126 Pisa, Italy}
\affiliation{International School for Advanced Studies (SISSA),
        via  Beirut 2-4,  I-34014 Trieste, Italy}

\date{\today}
\begin{abstract}
We present a diagrammatic real-time approach to adiabatic pumping of electrons
through interacting quantum dots.
Performing a systematic perturbation expansion in the tunnel-coupling strength,
we compute the charge pumped through a single-level quantum dot per pumping 
cycle.
The combination of Coulomb interaction and quantum fluctuations, accounted for 
in contributions of higher order in the tunnel coupling, modifies the pumping 
characteristics via an interaction-dependent renormalization of the 
quantum-dot level.
The latter is even responsible for the {\em dominant} contribution to the 
pumped charge when pumping via time-dependent tunnel-coupling strengths.
\end{abstract}

\maketitle

\section{Introduction}

In the absence of an applied bias voltage, a mesoscopic conductor can sustain 
a DC current component if two or more parameters of the device (for example 
gate voltages) are periodically modulated in time. 
Electron {\em pumping} is said 
to be \textit{adiabatic} if the parameter variation is slow on the scale 
defined by the dwell time of the electrons. In this case the pumped charge 
depends on the size and the shape of the pumping cycle but not on 
its detailed time evolution, i.e. it 
is of geometric nature.~\cite{geometric}
The idea of electron pumping dates back to a work of Thouless.~\cite{thouless} 
The first experiment on electron pumping in single electron devices was 
performed by Pothier \textit{et al.}~\cite{pothier} Since then much 
theoretical and experimental work has been devoted to electron 
pumping.~\cite{geometric,brouwer, zhou, levitov, levinson, 
aleiner1, moskalets, butti,mucciolo, spinpump,quantumhall,wangapl,blaauboer,
andreevinterf, switkes,dicarlo, watson,ahlers,green, sela, aleiner,citro,aono,
cota,brouwer2,shutenko,vavilov,polianski,woelfle}
For non-interacting systems the theory of adiabatic pumping, 
formulated by Brouwer,~\cite{brouwer} is based on the 
generalization~\cite{buttiker}  of 
the scattering approach for quantum transport to time-dependent phenomena. 
This formulation has been applied to study several aspects of pumping in 
non-interacting systems, such as the study of noise and 
decoherence,~\cite{levitov,moskalets,butti} the 
role of discrete symmetries,~\cite{aleiner1} the possibility of spin 
pumping,~\cite{mucciolo,spinpump,quantumhall} and the effect of 
superconducting elements and Andreev reflection.~\cite{wangapl,blaauboer,
andreevinterf} A diagrammatic approach was used in 
Refs.~\onlinecite{shutenko,vavilov,polianski} to calculate the pumped 
charge through a 
noninteracting system using random matrix theory in the limit of a large 
number of channels in the leads.
Furthermore several works investigate pumping by surface acoustic waves both 
theoretically~\cite{woelfle} and experimentally.~\cite{ahlers}
Pumping in interacting systems has been studied much less so far.
Quantum pumping through a Luttinger liquid has been discussed.~\cite{citro}  
In quantum dots pumping has been studied in the limit of weak 
interaction,~\cite{aleiner,brouwer2} in the Kondo regime,~\cite{aono} 
as well as in the Coulomb-blockade regime.~\cite{cota} 
In Refs.~\onlinecite{aleiner,brouwer2} the pumped charge through an open 
quantum dot is computed by means of the bosonization technique. Aono, 
in Ref.~\onlinecite{aono} uses
the Keldysh Green's function formalism complemented by the assumption
that the dot retarded Green's function takes the non interacting form
(this holds true in the non-interacting limit as well as in the Kondo regime).
The authors of Ref. ~\onlinecite{cota} integrate numerically the
master equation (in the Born-Markov approximation) for the reduced density
matrix of a double-dot pump.
Recently a general approach to pumping through 
interacting quantum dots has been put forward by relating the pumped charge to 
the instantaneous retarded Green's functions of the quantum 
dot.~\cite{green,sela}

In this paper we study adiabatic pumping through interacting quantum dots 
for temperatures much above the Kondo temperature but much below the level 
spacing in the dot. 
In this case a perturbative expansion in the tunnel coupling between the dot
and the leads is justified. 
Moreover we can restrict ourselves to consider only one level in the dot with 
a strong local repulsion in the case of double occupancy.
We aim at the understanding of the influence of Coulomb interaction on the 
pumping characteristics. 
In order to achieve this, we extend a diagrammatic real-time 
technique~\cite{koenig} that has been 
developed to describe non-equilibrium 
DC transport through an interacting quantum dot. 
As compared to the formalism in our recent work,~\cite{green} the perturbative 
approach presented here, although limited to weak tunnel-coupling strengths, 
is more transparent in identifying the physical origin of the various 
contributions to the pumped charge. 
In particular it is straightforward to relate the pumped current to the 
dynamics of the average charge of the dot. 

In this work we calculate the leading- and next-to-leading-order contribution 
of the perturbation expansion in the tunnel coupling 
to the pumped charge per pumping cycle. We distinguish two cases: pumping by 
changing periodically either the gate voltage and one tunnel barrier or 
varying both left and right tunnel barrier.
 Considering the first case, 
and furthermore taking into account only lowest-order tunneling processes 
associated with sequential tunneling, the adiabatic pump works analogously to a
peristaltic pump. The next-order correction turns out to be only due to a 
time-dependent
renormalization of the dot-level position that is induced by the combination
of Coulomb interaction and tunnel coupling to the leads. Remarkably this 
effect is not masked by other higher-order transport processes such as 
cotunneling.
The situation is even more dramatic for pumping with the tunnel barriers.
In this case, the lowest-order tunneling processes do not give rise to any
pumping. The dominant pumping mechanism is, then, of higher order, namely 
pumping by making use of the time-dependent level renormalization.
As a consequence, the gate-voltage dependence of the pumped charge provides a
transparent experimental access to probe quantum-fluctuation effects.

The paper is organized as follows: Section \ref{sec_model} introduces the 
model of the quantum dot. The time evolution of the system is 
described by a generalized Master equation in section \ref{sec_master} and an 
adiabatic expansion is performed and applied to the current pumped through the 
dot in section \ref{sec_current}. The expansion in the tunnel coupling 
is further discussed in section \ref{sec_tunnel}. 
The explicit evaluation of the formulae obtained up to here is done in 
section \ref{sec_rules} using a
diagrammatic technique. In section \ref{sec_results} results for the pumped 
current and the pumped charge are presented and discussed.

\section{Model and Formalism}
\subsection{Model\label{sec_model}}
We consider a single-level quantum dot with on-site Coulomb interaction 
coupled to two non-interacting leads. The system is described by the 
Hamiltonian 
$$
	H = H_{\text{leads}} + H_{\text{dot}} + H_{\text{tun}}
$$
where  $H_{\text{leads}}$, $H_{\text{dot}}$, and $H_{\text{tun}}$ describe, 
respectively,  the left (L) and right (R) leads,  the dot, and tunneling 
between dot and leads, and are given by
\begin{subequations}
\label{hamiltonians}
\begin{eqnarray}
	H_{\text{leads}} & = & \sum_{k,\sigma,\alpha} \epsilon_{k \alpha }   
	c_{\sigma k\alpha}^\dagger c_{\sigma k\alpha}
      \label{eqn_hamilton_leads}\\
	H_\text{dot} & = & \epsilon(t)  \sum_\sigma n_\sigma + U n_\uparrow 
	n_\downarrow\label{eqn_hamilton_dot}\\
	H_\text{tun} & = & \sum_{k,\sigma,\alpha} 
	\left[V_{\alpha}(t) c_{\sigma k\alpha}^\dagger d_\sigma+
          \mbox{h.c.}\right]
	\label{eqn_hamilton_tunnel}.
\end{eqnarray}
\end{subequations}
In Eqs.~(\ref{hamiltonians}), $c_{\sigma k\alpha}$ 
($c_{\sigma k\alpha}^\dagger $) 
is the fermionic  annihilation (creation) operator for an electron with spin 
$\sigma=\uparrow,\downarrow$,  momentum $k$, energy $\epsilon _k$ in lead 
$\alpha = \text{L},\text{R}$; $d_\sigma$ ($d_\sigma^\dagger$) 
is the fermionic annihilation (creation) operator for an electron with
spin $\sigma$ in the dot; and $n_\sigma=d^\dagger_\sigma d_\sigma$ is the 
number operator for the dot electrons with spin $\sigma$. The Coulomb 
interaction on the dot 
is described by the on-site energy $U$ associated with double occupation.
The leads are assumed to be in thermal equilibrium with the same chemical 
potential and to have flat bands with constant density of states $\rho_\alpha$.

By periodically changing at least two of the system parameters, a DC current 
can be pumped through the dot. We choose the level position of the dot
$\epsilon(t) $ 
and the  tunnel matrix elements $V_{\alpha}(t)$ to be time-dependent.
We only allow for the modulus, but not the phase, of $V_{\alpha}(t)$ to vary in
time, since a time-dependent phase would correspond to a bias voltage. 
We define the time-dependent intrinsic line width 
$\Gamma_{\alpha}\left(t,t'\right)=2\pi\rho_{\alpha}V_{\alpha}(t)
V^{*}_{\alpha}(t')$, the total intrinsic line width $\Gamma\left(t,t'\right)
=\Gamma_{\text{L}}\left(t,t'\right)+\Gamma_{\text{R}}\left(t,t'\right)$, as 
well as $\Gamma_{\alpha}\left(t\right)=\Gamma_{\alpha}\left(t,t\right)$ and
$\Gamma \left(t\right)=\Gamma \left(t,t\right)$.
To keep all formulae transparent, we set $\hbar \equiv 1$ throughout the paper.

\subsection{Generalized Master equation and adiabatic approximation}
\label{sec_master} 

As described above we consider an interacting quantum dot coupled to 
non-interacting leads. Since the leads act as baths, it is convenient
to trace out the degrees of freedom of the non-interacting lead states to 
arrive at an effective description of the reduced system. In the limit of 
temperature much lower than the level spacing of the dot, only one 
level will contribute to transport. Therefore the Hilbert space for the dot 
is four dimensional: the quantum dot can be empty, singly occupied with a 
spin-up or a spin-down electron or doubly occupied. These states, labeled 
by $\chi = \left\{ 0,\ \uparrow,\ \downarrow,\ \text{d} \right\}$, have energy 
$E_0=0$, $E_\uparrow =E_\downarrow=\epsilon$ and $E_\text{d}=2\epsilon +U$, 
respectively. In the following we use a matrix notation 
in the four dimensional Hilbert space of the dot, with boldface symbols  for 
vectors 
and matrices. The probabilities to find the dot in the respective state are
$\mathbf{p}=  (p_0,p_\uparrow,p_\downarrow,p_{\text{d}})^{\text{T}}$.

The starting point of our analysis is the generalized Master equation for the
time evolution of the system,
\begin{equation}
\label{eqn_determine_p}
	\frac{d}{dt}\mathbf{p}\left(t\right)=\int_{-\infty}^{t}dt'\mathbf{W}
	\left(t,t'\right)\mathbf{p}\left(t'\right)\, ,
\end{equation}
where the matrix elements $W_{\chi,\chi'}\left(t,t'\right)$ of the kernel 
$\mathbf{W}\left(t,t'\right)$ describe the transition from a state $\chi'$ at 
time $t'$ to a state $\chi$ at time $t$.
For the system considered here, Eq. (2) defines the most general kinetic 
equation for the dot probabilities without any approximation.
Off-diagonal matrix elements of the reduced density matrix for the quantum dot,
that correspond to {\it real} superposition of different states 
$\chi \neq \chi'$, do not couple to the diagonal ones since $\chi$ and $\chi'$ 
differ by a conserved quantum number, particle number or spin.
Therefore, off-diagonal elements of the reduced density matrix
do not enter any transport quantity.
Nevertheless, quantum fluctuation effects involving {\it virtual} intermediate
states of the quantum dot in higher-order processes such as cotunneling, are
fully taken into account by Eq. (2) by properly evaluating the kernel $\mathbf{W}$.~\cite{comment}

Our goal is to describe the response of the system to slow periodic variations 
of the system parameters $X(t)$ with frequency $\Omega$.
After waiting long enough, such that any memory of the initial dot-state
distribution $\mathbf{p}\left(-\infty\right)$ has died out, the dynamics of 
the system is fully determined by the explicit time dependence of the system 
parameters.
The latter enters Eq.~(\ref{eqn_determine_p}) in two ways, namely by the
kernel $\mathbf{W}\left(t,t'\right)$ being a functional of the system 
parameters $X(\tau)$ with $\tau \in [t',t]$, and by the non-Markovian 
structure, i.e., the time derivative of $\mathbf{p}\left(t\right)$ at time 
$t$ depends on $\mathbf{p}\left(t'\right)$ at an earlier time $t'$ at which 
the system parameters had different values.
In the adiabatic regime it is possible to simplify considerably the form of the
Master equation by performing an {\em adiabatic expansion}, i.e., an
expansion in the pumping frequency $\Omega$, assuming that $\Omega$ is small 
as compared to both the energy scales that determine the decay time of the 
kernel and the time integral of the kernel (which sets the time scale of
the system's response to the parameter's change).
The zeroth-order, \textit{instantaneous}, term of the adiabatic expansion
corresponds to freezing the value of all system parameters, $X(t)$, at time 
$t$, which corresponds to solving a time-independent problem.
To obtain the first-order correction we need to systematically collect all
contributions linear in the pumping frequency or, equivalently, linear in
the time derivative of $X$ at time $t$.
For this, we perform a Taylor expansion of $\mathbf{p}(t')$ around $t$ up to
linear order,
\begin{equation}
\label{master_appr}
	\frac{d}{dt}\mathbf{p}\left(t\right)=\int_{-\infty}^{t}dt'\mathbf{W}
	\left(t,t'\right)\left[\mathbf{p}\left(t\right)+(t'-t)\frac{d}{dt}
         \mathbf{p}(t)\right]\, .
\end{equation}

Furthermore, we perform an adiabatic expansion of the kernel 
$\mathbf{W}\left(t,t'\right)$ itself.
The zeroth-order term, $\mathbf{W}^{(i)}_{t}(t-t')$, is indicated with the 
superscript $(i)$ for {\em instantaneous} and the subscript $t$ to emphasize 
that the system parameters $X(\tau) \rightarrow X(t)$ are frozen at time $t$.
It depends only on the time difference $t-t'$, and only parametrically on 
$t$ through $X(t)$.
The first-order term is obtained by linearizing the time dependence of all 
parameters $X(\tau)$ with respect to the final time $t$, i.e., 
$X(\tau) \rightarrow X(t)+ (\tau-t) \frac{d}{d\tau} X(\tau)|_{\tau=t}$, and 
retaining only linear terms in time derivatives.
This linear correction to the kernel is indicated by 
the superscript $(a)$ for {\em adiabatic},
\begin{equation} 
\label{W_exp}
  	\mathbf{W}(t,t') \rightarrow \mathbf{W}^{(i)}_{t}(t-t') + 
  	\mathbf{W}^{(a)}_{t}(t-t') \, .
\end{equation}

Finally, we need to perform an adiabatic expansion for the occupation 
probabilities in the dot,
\begin{equation}
\label{p_exp}
	\mathbf{p}(t)\rightarrow \mathbf{p}^{(i)}_{t}+\mathbf{p}^{(a)}_{t}.
\end{equation} 
The instantaneous probabilities $\mathbf{p}^{(i)}_{t}$ are the solution 
of the time-independent problem with all parameter values fixed at time $t$.
They are obtained from the Master Eq.~(\ref{master_appr}) in the stationary 
limit,
\begin{equation}
  	0  =  \mathbf{W}^{(i)}_{t}\mathbf{p}^{(i)}_{t} \, ,
  	\label{eqn_instantaneous_P}
\end{equation}
together with the normalization condition 
$\mathbf{e}^{\text{T}}\mathbf{p}^{(i)}_{t}=1$, where 
$\mathbf{e}=(1,1,1,1)^{\text{T}}$, and we have introduced the Laplace 
transform 
\begin{equation}\nonumber
	F(z) = \int_{-\infty}^{t}dt' e^{-z(t-t')} F(t-t')
\end{equation}
to define $\mathbf{W}^{(i)}_{t}= \mathbf{W}^{(i)}_{t}(z=0_+)$.
The first adiabatic correction can be obtained from Eq.~(\ref{master_appr}), 
using Eqs.~(\ref{W_exp}) and (\ref{p_exp}).
We find
\begin{equation}
	\mathbf{W}^{(i)}_t \mathbf{p}^{(a)}_t=
	 \frac{d}{dt}\mathbf{p}^{(i)}_t -
  	\mathbf{W}^{(a)}_t \mathbf{p}^{(i)}_t -
   	\partial \mathbf{W}^{(i)}_t \frac{d\mathbf{p}^{(i)}_t}{dt}  
  	\, ,
\label{eqn_ad_expansion_P} 
\end{equation}
where again $\mathbf{W}^{(i/a)}_{t}$ is the Laplace transform at zero 
frequency and
$\partial \mathbf{W}^{(i)}_{t} = (\partial / \partial z)
\mathbf{W}^{(i)}_{t}(z) |_{z = 0_+}$.
Once $\mathbf{W}^{(i/a)}_{t}$ are evaluated and the instantaneous
probabilities $\mathbf{p}^{(i)}_t$ are known from 
Eq.~(\ref{eqn_instantaneous_P}), the adiabatic corrections 
$\mathbf{p}^{(a)}_t$ are
obtained from Eq.~(\ref{eqn_ad_expansion_P}) together with the normalization
condition $\mathbf{e}^{\text{T}}\mathbf{p}^{(a)}_{t} = 0$.

\subsection{Pumped charge}\label{sec_current}

The charge $Q$ pumped in one cycle $\mathcal{T}=1/\Omega$ is related to 
the time-dependent current $I_{\text L}(t)$ flowing through the left barrier 
by 
\begin{equation}\nonumber
	Q=\int_0^{\mathcal{T}} I_{\text{L}}(t) dt \,\,\,\, .
\end{equation}
By accounting for the time evolution of the system before time $t$ in a similar
way as done for the dot-state probabilities above, we can express the current 
into the left lead as
\begin{eqnarray}
\label{eqn_current}
	I_{\text{L}}\left(t\right)=e\int_{-\infty}^{t}dt'\mathbf{e}^{\text{T}}
	\mathbf{W}^{\text{L}}\left(t,t'\right)\mathbf{p}\left(t'\right),
\end{eqnarray}
where $\mathbf{W}^{\text{L}}\left(t,t'\right)=\sum_p p
\mathbf{W}^{\text{L} p} \left(t,t'\right)$, and 
$\mathbf{W}^{\alpha p}\left(t,t'\right)$ includes all
processes associated with transitions where the number of electrons
(with charge $e$) entering reservoir $\alpha$ minus the ones leaving it 
equals $p$.

It is straightforward to perform an adiabatic expansion for 
Eq.~(\ref{eqn_current}) in the same way as for the Master equation.
The instantaneous or zeroth-order level of the adiabatic expansion is 
sufficient to describe the DC current that is driven through the quantum dot
by an applied transport voltage.~\cite{koenig}
It is, furthermore, sufficient for modeling {\em rectification}, i.e., the
generation of a DC current component by applying an AC transport voltage
and appropriately changing some system parameter in time.
In the absence of any DC or AC transport voltage, as considered in the present
paper, the instantaneous part of the current vanishes. In order to describe 
{\em pumping}, one needs to compute the first-order adiabatic correction of 
the current. 
Using for Eq.~(\ref{eqn_current}) the same procedure as for the Master 
equation, we find the adiabatic part of the current to be 
\begin{equation}
\label{eqn_current_ad}
  	I_{\text{L}}\left(t\right) = e \ \mathbf{e}^{\text{T}} \left[
  	\mathbf{W}^{\text{L}(a)}_t \mathbf{p}^{(i)}_t
  	+ \mathbf{W}^{\text{L}(i)}_t \mathbf{p}^{(a)}_t
  	+ \partial \mathbf{W}^{\text{L}(i)}_t 
  	\frac{d \mathbf{p}^{\left(i\right)}_t}{dt}
  	\right] .
\end{equation}

\subsection{Perturbation expansion in tunneling}\label{sec_tunnel}

Alongside with the adiabatic expansion we perform a perturbation expansion in 
powers of the tunnel coupling strength $\Gamma$ for both the instantaneous 
and the adiabatic correction of the kernel $\mathbf{W}$, the probabilities
$\mathbf{p}$, and the current $I_{\text L} (t)$.
We indicate the order of the perturbation expansion in $\Gamma$ by adding a 
superscript, i.e., $\mathbf{W}^{(i)}_t = \mathbf{W}^{(i,1)}_t +
\mathbf{W}^{(i,2)}_t + {\cal O}(\Gamma^3)$ for the instantaneous contribution
to the kernel, and similarly for $\mathbf{W}^{(a)}_t$.
The expansion of the instantaneous probabilities begins in zeroth order
in $\Gamma$, $\mathbf{p}^{(i)}_t = \mathbf{p}^{(i,0)}_t + \mathbf{p}^{(i,1)}_t 
+ {\cal O}(\Gamma^2)$, in order to be able to fulfill the normalization 
condition $\mathbf{e}^{\text{T}} \mathbf{p}^{(i)}_t = 1$.
By expanding Eq.~(\ref{eqn_instantaneous_P}) in powers of $\Gamma$, we find 
that the instantaneous probabilities should fulfill the two equations,
\begin{subequations}
\label{instprob}
\begin{eqnarray}
  	0 & = & \mathbf{W}^{(i,1)}_t \mathbf{p}^{(i,0)}_t\\
  	0 & = & \mathbf{W}^{(i,2)}_t \mathbf{p}^{(i,0)}_t 
  	+ \mathbf{W}^{(i,1)}_t \mathbf{p}^{(i,1)}_t \, ,
\end{eqnarray}
\end{subequations}
together with the normalization conditions 
$\mathbf{e}^{\text{T}} \mathbf{p}^{(i,0)}_t = 1$ and 
$\mathbf{e}^{\text{T}} \mathbf{p}^{(i,1)}_t = 0$.
As discussed above, the instantaneous part of the current vanishes in all
order in $\Gamma$ due to the absence of an applied transport voltage.

In order to determine the adiabatic corrections to the probabilities, we 
expand also Eq.~(\ref{eqn_ad_expansion_P}) in powers of $\Gamma$,
\begin{subequations}
\label{adprob}
\begin{eqnarray}
  \label{adprob_a}
  \frac{d\mathbf{p}^{(i,0)}_t}{dt} & = & \mathbf{W}^{(i,1)}_t
  \mathbf{p}^{(a,-1)}_t\\
  \frac{d\mathbf{p}^{(i,1)}_t}{dt} & = & \mathbf{W}^{(i,1)}_t
  \mathbf{p}^{(a,0)}_t + \mathbf{W}^{(i,2)}_t \mathbf{p}^{(a,-1)}_t
  \nonumber\\
  & & + \mathbf{W}^{(a,1)}_t\mathbf{p}^{(i,0)}_t
  + \partial \mathbf{W}^{(i,1)}_t \frac{d\mathbf{p}_t^{(i,0)}}{dt}
    \, ,
\end{eqnarray}
\end{subequations}
with the normalization conditions 
$\mathbf{e}^{\text{T}} \mathbf{p}^{(a,-1)}_t = 0$ and 
$\mathbf{e}^{\text{T}} \mathbf{p}^{(a,0)}_t = 0$.
We emphasize that, in order to properly match the powers of $\Gamma$ on the 
left and right hand side of Eq.~(\ref{adprob}), one has to start the expansion
of the adiabatic correction of the probabilities in {\em minus first} order in 
$\Gamma$, $\mathbf{p}^{(a)}_t = \mathbf{p}^{(a,-1)}_t + \mathbf{p}^{(a,0)}_t 
+ {\cal O}(\Gamma)$.
At first glance, such an expansion might look divergent for the weak-coupling 
limit, $\Gamma \rightarrow 0$.
However, in the validity range of the adiabatic expansion everything remains
well defined: the adiabaticity condition requires that the energy scale defined
by the pumping frequency $\Omega$ is much smaller than the tunnel-coupling 
strength $\Gamma$. 
Since the time derivative on the left hand side of Eq.~(\ref{adprob}) 
introduces a factor $\Omega$, we see that $\mathbf{p}^{(a,-1)}_t$ scales with
$\Omega/\Gamma$, which is always much smaller than $1$
in the adiabatic limit. 

The perturbation expansion of the adiabatically pumped current is 
derived from Eq.~(\ref{eqn_current_ad}) and reads
\begin{subequations}
\label{expcur}
\begin{eqnarray}
  	I_{\text{L}}^{(0)} (t) &=& e\ \mathbf{e}^{\text{T}}
  	\mathbf{W}^{\text{L}(i,1)}_t \mathbf{p}^{(a,-1)}_t\\
  	I_{\text{L}}^{(1)} (t) &=& e\ \mathbf{e}^{\text{T}} \left[
  	\mathbf{W}^{\text{L}(i,1)}_t \mathbf{p}^{(a,0)}_t
  	+ \mathbf{W}^{\text{L}(i,2)}_t \mathbf{p}^{(a,-1)}_t \right.
  	\nonumber \\
  	&& \left. + \mathbf{W}^{\text{L}(a,1)}_t \mathbf{p}^{(i,0)}_t
  	+ \partial \mathbf{W}^{\text{L}(i,1)}_t \frac{d\mathbf{p}_t^{(i,0)}}{dt}
  	\right] \, .
\end{eqnarray}
\end{subequations}  
We see that the lowest-order contribution to the pumped current starts in 
zeroth order in the tunnel coupling strength $\Gamma$, as it consists in a 
product of a first- and a minus-first order term in the tunneling coupling,
but scales linearly with the pumping frequency $\Omega$.
This contrasts with the DC current driven by a finite bias voltage, for which 
the lowest-order contribution is linear in $\Gamma$.

Certain properties of pumping can be derived by a closer inspection of the 
perturbative expansion of the Master equation.
To zeroth order in $\Gamma$ the pumped current is nonzero only if 
$\mathbf{p}^{(a,-1)}$ 
is non vanishing, which, according to Eq.~(\ref{adprob_a}) requires that the 
zeroth-order instantaneous probabilities $\mathbf{p}^{(i,0)}$ depends on time.
However the latter are simply determined by the Boltzmann factors of the 
corresponding state energies:
\begin{equation}
p^{(i,0)}_\chi=\frac{e^{-\beta E_{\chi}}}{Z} \ ,\nonumber
\end{equation}
where $E_{\chi}$ is the energy related to the dot state 
$\chi$, $\beta=1/k_{\mathrm{B}}T$ is the inverse temperature and $Z$ the 
partition function.
In particular, the probabilities $p^{(i,0)}_\chi$ are independent of the 
tunnel couplings. As a consequence, in order to have a non-vanishing 
zeroth-order pumped current $I_{\text{L}}^{(0)}$, one of the pumping 
parameters has to be the level position.
When pumping with the two barrier heights, $I_{\text{L}}^{(0)}$ vanishes.

\subsection{Diagrammatic rules}\label{sec_rules}

In order to evaluate the kernel ${\bf W}$ of the Master equation 
Eq.~(\ref{eqn_determine_p}) we use  
the diagrammatic perturbation approach to transport through interacting 
quantum dots developed in Ref.~\onlinecite{koenig}.
While in Ref.~\onlinecite{koenig} the diagrammatic language was derived for 
DC transport with time-independent system parameters, we generalize the 
approach in this section to account for the adiabatic time dependence of the 
external parameters.

We start with deriving the Master equation from a very general point 
of view in order to relate its  kernel  to a set of diagrams to be evaluated.
In general, the (time-dependent) transport properties are governed by the 
time evolution of the reduced density matrix of the dot obtained after
tracing out the degrees of freedom of the non-interacting lead electrons. 
Since the leads are non-interacting and in thermal equilibrium, the lead 
electrons can be  integrated 
out making use of Wick's theorem, i.e. contracting pairs of 
creation and annihilation operators 
$c_{\sigma k\alpha}^\dagger$ ($c_{\sigma k \alpha}$). 
Furthermore, since in the case of pumping there is no voltage applied between 
left and right lead, 
the occupation of electronic states in both leads is described by the same 
Fermi distribution function $f(\omega)$.
The time evolution of the reduced density matrix is related to the propagator
$\mathbf{\Pi}\left(t,t'\right)$ by
\begin{equation}\label{eqn_propagator}
\mathbf{p}\left(t\right)=\mathbf{\Pi}\left(t,t'\right)
\mathbf{p}\left(t'\right) \, .
\end{equation}

\begin{figure}
\includegraphics[width=2.5in]{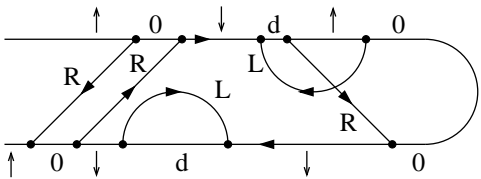}
\caption{Example for the time evolution of the reduced density matrix. The 
upper and lower line represent the forward and backward time propagation along 
the Keldysh contour. Tunneling lines connecting vertices represent tunneling 
events with the left (right) reservoir. Next to the propagators the respective 
dot states are indicated. 
\label{fig_red_density}}
\end{figure}

\begin{figure}
\includegraphics[width=2.7in]{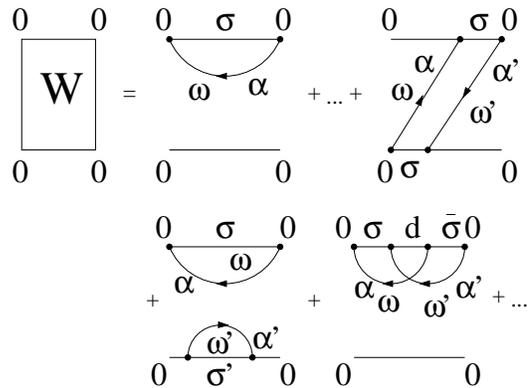}
\caption{Examples of diagrams contributing to the irreducible diagram part 
$W_{00}$.
\label{fig_block_example}}
\end{figure}

Contributions to this propagator can be depicted as diagrams on the Keldysh 
contour, where contractions of fermion operators of the leads are indicated 
as tunneling lines.
An example is shown in Fig.~\ref{fig_red_density}.
The propagator $\mathbf{\Pi}\left(t,t'\right)$ can be expressed in terms of 
its irreducible part $\mathbf{W}\left(t''',t''\right)$ by means of a Dyson 
equation:
\begin{equation}\label{eqn_reduced_self}
\mathbf{\Pi}\left(t,t'\right)=\mathbf{1}+\int_{t'}^{t}dt'''\int_{t'}^{t'''}dt''
\mathbf{W}\left(t''',t''\right)\mathbf{\Pi}\left(t'',t'\right) \, .
\end{equation}
The irreducible diagram part $\mathbf{W}\left(t''',t''\right)$ is
 defined as the sum over all diagrams in which 
any vertical cut crosses at least one tunneling line (see 
Fig.~\ref{fig_block_example} as an example).\\
Performing the time derivative of Eq.~(\ref{eqn_propagator}),
plugging in Eq.~(\ref{eqn_reduced_self}), and shifting the lower bound of the
remaining time integral to minus infinity, we obtain the generalized Master 
equation of Eq.~(\ref{eqn_determine_p}).

In a similar way we proceed for the current $I_{\alpha}\left(t\right)=
e\frac{d}{dt}\langle N_{\alpha}\left(t\right)\rangle$ of particles flowing 
into reservoir $\alpha$, which is given by
\begin{equation}
I_{\alpha}\left(t\right)=-
ie\sum_{k,\sigma}\left[V_{\alpha}\langle c^{\dagger}_{\sigma k\alpha}d_{\sigma}
\rangle (t) - V^{*}_{\alpha}\langle d_{\sigma}^{\dagger}c_{\sigma k\alpha}
\rangle (t) \right] \, .
\end{equation}
It can also be expressed in terms of diagrams. They contain a vertex at time 
$t$ that has the same structure as the tunneling vertices in 
Fig.~\ref{fig_red_density}, and can be attached to 
the upper or the lower propagator line.
We find Eq.~(\ref{eqn_current}) where $\mathbf{W}^{\alpha}\left(t,t'\right)
=\sum_p p \mathbf{W}^{\alpha p}\left(t,t'\right)$ has the 
following properties. The terms $\mathbf{W}^{\alpha p}\left(t,t'\right)$ are 
given by all diagrams for which 
the number of tunneling lines with reservoir index $\alpha$ running from the 
upper to the lower propagator 
minus the number of those with reservoir index $\alpha$ running in the opposite
direction equals $p$.

In the following we summarize the diagrammatic rules for the kernel 
$\mathbf{W}_t^{(i)}$ (as described in Ref.~\onlinecite{koenig}) and discuss 
additional rules for the evaluation of its adiabatic expansion. 
Examples for the application of the rules given below are shown in 
appendix~\ref{ap_example}.
We start with rules for the Laplace transform of $\mathbf{W}_t^{(i,n)}(z)=
\int_{-\infty}^{t}dt' \exp(-z(t-t')) \mathbf{W}^{(i,n)}_t \left(t,t'\right)$
in $n$-th order in the tunnel coupling.
This will directly lead us to the desired objects
$\mathbf{W}_t^{(i,n)} = \mathbf{W}_t^{(i,n)}(z)|_{z=0_+}$ and
$\partial \mathbf{W}_t^{(i,n)} = (\partial/\partial z) 
\mathbf{W}_t^{(i,n)}(z)|_{z=0_+}$.
\\
 (1) Draw all topologically different diagrams with $n$ directed tunneling 
lines connecting pairs of vertices containing lead electron operators. 
Assign a reservoir index $\alpha$, an energy $\omega$ and a spin index 
$\sigma$ to each of these lines. 
Assign states $\chi$ and the corresponding energies $E_{\chi}\left(t\right)$ 
to each element of the Keldysh contour connecting two vertices.
Furthermore, draw an external line from the upper leftmost beginning of a dot 
propagator to the upper rightmost end of a dot propagator that carries the 
(imaginary) energy $-iz$.\\
 (2) For each time segment between two adjacent vertices (independent on 
whether they are on the same or on opposite branches of the Keldysh contour)
write a resolvent 
$1/ \Delta E\left(t\right)$ where $\Delta E\left(t\right)$ 
is the difference of left going minus right going energies (including energies 
of tunneling lines and the external line - the positive imaginary part of
$iz$ will keep all resolvents regularized).\\
 (3) Each vertex containing a dot operator $d_{\sigma}^{(\dagger)}$ gives rise 
to a matrix element $\langle\chi'|d_{\sigma}^{(\dagger)}|\chi\rangle$ where 
$\chi$ ($\chi'$) is the dot state entering (leaving) the vertex with respect 
to the Keldysh contour.\\
 (4) The contribution of a tunneling line of reservoir $\alpha$ is 
$\frac{1}{2\pi}\Gamma_{\alpha}\left(t\right)f\left(\omega\right)$ if the line 
is going backward with respect to the closed time path and  
$\frac{1}{2\pi}\Gamma_{\alpha}\left(t\right)
\left[1-f\left(\omega\right)\right]$ if it is going forward.\\
 (5) The overall prefactor is given by 
$\left(-i\right)\left(-1\right)^{b+c}$ where $b$ is the total 
number of vertices on the backward propagator and $c$ the number 
of crossings of tunneling lines.\\
 (6) Integrate over the energies of tunneling lines and sum over reservoir and 
spin indices.\\

\begin{figure}
\includegraphics[width=2.5in]{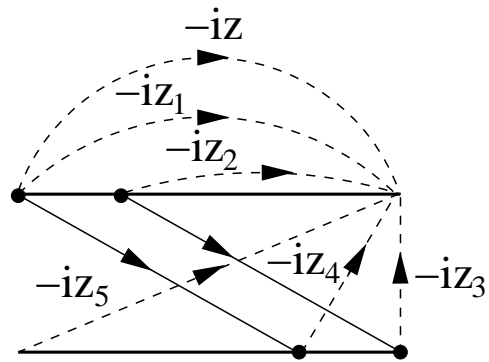}
\caption{Example on how to add the external frequency lines. 
The line $-iz$ is needed for the evaluation of $\partial \mathbf{W}$; 
$-iz_1$, $-iz_2$, and $-iz_4$ are needed for 
the contributions to $\mathbf{W}^{(\text{a})}$ due to 
both $\frac{d\Gamma}{dt}$ and 
$\frac{d\epsilon}{dt}$; $-iz_3$ does not contribute and can be omitted; 
and $ -iz_5$ is additionally needed for the evaluation of the contribution to 
$\mathbf{W}^{(\text{a})}$ due 
to  $\frac{d\epsilon}{dt}$.}
\label{adiabaticdiagrams}
\end{figure}

To derive the rules for the adiabatic corrections $\mathbf{W}_t^{(a,n)}$
we first analyze how the time-dependent parameters enter the 
expression of the kernel $\mathbf{W}(t,t')$.
The time-dependent variables for which the adiabatic expansion has to be 
performed are $V_{\alpha}\left(t\right)$ and $\epsilon\left(t\right)$, where 
the first one only appears in the product $\Gamma_{\alpha}\left(t_i,t_j\right)
=2\pi\rho V_{\alpha}\left(t_i\right)V_{\alpha}\left(t_j\right)$ associated with
a tunneling line, and the latter only in the isolated-dot propagator 
$\exp\left(- i\int_{t_i}^{t_j}dt' E_{\chi}\left(t'\right)\right)$ for each 
segment between adjacent vertices. While for the instantaneous kernels all 
parameters were taken at time $t$, now we perform a series expansion around 
the same time $t$ and keep all contributions linear in a time derivative of
the pumping parameters,
\begin{eqnarray}
\Gamma\left(t_i,t_j\right) & \approx &  \Gamma\left(t\right)
\label{firstorder_gamma_arb}\\
 &+&\frac{t_i-t}{2}\frac{d\Gamma}{dt}\left(t\right)+
\frac{t_j-t}{2}\frac{d\Gamma}{dt}\left(t\right)
\nonumber\\
e^{- i\int\limits_{t_i}^{t_j}dt' E_{\chi}(t')} & \approx & e^{- i 
E_{\chi}\left(t\right)\cdot \left(t_j-t_i\right)} \times
\nonumber \\ && \!\!\!\!
\left[1 - i\frac{\left(t_j-t\right)^{2} - \left(t_i-t\right)^{2}}{2}
\frac{d E_{\chi}}{dt}\left(t\right) \right] .  
\label{firstorder_eps_arb}
\end{eqnarray}
The factors $(t_i-t)$ or $(t_i-t)^2$ can be included in the diagrammatic rules
in the following way: introduce an additional external frequency line with 
the imaginary energy $-i z_i$ from the vertex at $t_i$ to the rightmost vertex 
at $t$ (or the imaginary energy $-i z_j$ from the beginning of a dot 
propagator line at $t_j$ to the rightmost upper end of a dot 
propagator line at $t$), performing the first derivative with respect to 
$z_i$ (or second derivative with respect to $z_j$)
then set $z_i = 0_+$ and $z_j = 0_+$.
The external frequency lines are drawn as dotted lines in 
Fig.~\ref{adiabaticdiagrams}.

The rules to  compute the contribution to the adiabatic corrections 
$\mathbf{W}_t^{(a,n)}$ due to the time-dependence of $\Gamma(t)$ read:\\
(7a) Add to all diagrams needed for $\mathbf{W}_t^{(i,n)}(z)$ 
additional external frequency lines between any vertex $t_i$ and the right 
corner of the diagram and assign to them an (imaginary) energy $-iz_i$. 
Note that an eventual external frequency line between two right corners of a 
diagram does not contribute and can always be omitted.\\
(7b) Follow the rules (1) to (6) taking into account the extra lines.\\
(7c) Perform a first derivative with respect to 
$z_i$ and multiply it by the factor 
$\frac{1}{2}\frac{d\Gamma}{dt}\left(t\right)\frac{1}{\Gamma\left( t\right)}$. 
Sum all the contributions obtained in this way.\\ 
(7d) Set all the external frequencies $z_i$ and $z$ to $0_+$.

The contribution to the adiabatic correction  $\mathbf{W}_t^{(a,n)}$ due to 
the time-dependence of the level position can be computed in a similar way:  \\
(8a) In addition to the external frequency lines added according to rule (7a), 
put one more external frequency line from the left corner of the diagram 
with no vertex to the right corner.  \\
(8b) Follow the rules (1) to (6) taking into account the extra lines.\\  
(8c) Perform a second derivative with respect to $z_i$
and multiply by $-\frac{i}{2}\frac{d(E_{\chi}-E_{\chi'})}{dt}\left(t\right)$, 
where $\chi$ ($\chi'$) is the dot state entering (leaving) the vertex of the 
external frequency line at $t_i$ with 
respect to the Keldysh contour. The term  $\frac{d E_{\chi}}{dt}$ 
($\frac{d E_{\chi'}}{dt}$) is omitted if the segment associated 
with $E_{\chi}$ ($E_{\chi'}$) does not belong to the diagram. Sum all the 
contributions obtained in this way. 
\\ 
(8d) Set the external frequencies $z_i$ and $z$ to $0_+$.

\section{Results\label{sec_results}}

In this section we show the results for the pumped current and the pumped 
charge 
through a single-level quantum dot. As no bias voltage is 
applied, the only 
contribution to the pumped current arises from the adiabatic correction, 
and  hence we drop the superscript $(a)$ for the current. 
As we discuss below, the properties of the pumped charge, in the regime 
discussed in this paper, can be understood to a large extent in terms of 
the time-dependence of the occupation of the quantum dot, 
$\langle n\rangle=p_\uparrow+p_\downarrow+2 p_\text{d}$. 
For this reason, we first want to discuss the perturbation expansion of the
instantaneous average charge occupation.
The contribution to zeroth order in $\Gamma$ turns out to be determined by 
the Boltzmann factors of the energies associated with the states of the 
isolated quantum dot.
This yields
\begin{equation}
  	\langle n\rangle^{(i,0)}=
          \frac{2 f(\epsilon)}{1+f\left(\epsilon\right)-
    	f\left(\epsilon+U\right)}. 
\end{equation}
The first-order correction accounts for quantum fluctuations due to tunneling
from and to the leads.
There are two qualitatively different effects which are due to tunneling
and correspondingly we present the results for the first-order corrections as 
a sum of two contributions $\langle n\rangle^{(i,1)} = 
\langle n \rangle^{(i,\text{broad})} + \langle n \rangle^{(i,\text{ren})} $, 
where the contribution to broadening is the sum of the contributions of the two
 leads $\langle n \rangle^{(i,\text{broad})} = \langle n \rangle
^{(i,\text{broad,R})} + \langle n \rangle^{(i,\text{broad,L})}$.

First, the dot levels acquire a finite life-time broadening due to the coupling
to lead $\alpha$, which is accounted for by 
\begin{equation}
\label{eqn_n_broad}
	\langle n \rangle^{(i,\text{broad})} =
	\left(2-\langle n\rangle^{(i,0)}\right)	\phi'\left(\epsilon\right)
	+ \langle n\rangle^{(i,0)}\phi' \left(\epsilon+U\right),
\end{equation}
where $\phi'(\omega)$ is the derivative of 
$\phi\left(\omega\right)=\frac{\Gamma}{2 \pi} \mathbb{R}\mathrm{e}\Psi 
\left(\frac{1}{2}+ \frac{i \beta \omega}{2 \pi}\right)$, and $\Psi$ is the 
digamma function. The first term accounts for the broadening of the resonance
at $\epsilon$ between empty and singly-occupied dot; it has a prefactor $2$, 
when the dot is empty, and is zero when the dot is doubly occupied. 
The second term accounts for the broadening of the resonance at $\epsilon+U$
between singly- and doubly-occupied dot (this contribution is zero if the dot 
is empty). 
In the case that the dot is singly occupied both terms contribute with a 
prefactor $1$. 

Second, the combination of tunneling and charging energy gives 
rise to a renormalization of the level position, 
$\epsilon  \rightarrow  \epsilon+ \sigma\left(\epsilon,\Gamma,U\right)$ with 
\begin{equation}
\label{eqn_renormalization}
	\sigma\left(\epsilon,\Gamma,U\right)=\phi\left(\epsilon+U\right)-
	\phi\left(\epsilon\right)\, ,
\end{equation} 
as it is expected from the poor man's scaling analysis.~\cite{poorman}
The level renormalization is positive when the level is in the vicinity of 
the Fermi energy of the leads and negative for $\epsilon+U$ being close 
to the Fermi energy. This means that the distance between the two resonances 
from empty to singly-occupied dot and from singly- to doubly-occupied dot is 
effectively decreased.
The changes to the instantaneous average occupation due to the level 
renormalization reads
\begin{equation}
\label{eqn_n_renormalization}
  	\langle n \rangle^{(i,\text{ren})}=\frac{d}{d \epsilon}\left(\langle 
  	n\rangle^{(i,0)} \right)       \sigma\left(\epsilon,\Gamma,U\right) 
          \, .
\end{equation}
The sum of Eqs.~(\ref{eqn_n_broad}) and (\ref{eqn_n_renormalization}) is 
directly found from the correction in first order $\Gamma$ to the occupation 
probability, which is shown explicitly in appendix~\ref{ap_results}.
We remark that the above interpretation of the two terms is in full agreement
with known exact results for the non-interacting case.
For $U=0$ (and flat density of states in the leads), the level renormalization 
vanishes.
The spectral density is then equal to the Breit-Wigner function and its 
expansion in zeroth and first order in $\Gamma$ leads to the non-interacting 
result of $\langle n \rangle^{(i,0)}$ and 
$\langle n \rangle^{(i,\text{broad})}$.

\subsection{Adiabatically-pumped current}

We now proceed with solving Eqs.~(\ref{instprob}a), (\ref{adprob}a) and 
(\ref{expcur}a) for arbitrary interaction $U$ to get the zeroth-order 
adiabatically pumped current.
The result of  the diagrammatic approach explained above 
can be written in the form
\begin{eqnarray}
  	\label{eq_first_current}
  	I_{\text{L}}^{(0)}\left(t\right) & = & 
              -e\frac{\Gamma_{\text{L}}}{\Gamma}
  	\frac{d}{dt}\langle n\rangle^{(i,0)} \, . 
\end{eqnarray}
This suggests the following interpretation.
As the dot occupation is changed in time by varying the pumping parameters
(one of them must be the level position since $\langle n\rangle^{(i,0)}$ is 
independent of the tunnel-coupling strengths), the charge moves in and out of  
the quantum dot generating a current from/into the leads.
The contributions flowing through barrier $\alpha$ split weighted by the
time-dependent relative tunnel couplings $\Gamma_{\alpha}/\Gamma$.

By means of Eqs.~(\ref{instprob}b), (\ref{adprob}b) and (\ref{expcur}b), one 
finds the first-order-$\Gamma$ contribution to the current,
\begin{equation}
	I_{\text{L}}^{(1)}\left(t\right) =  -e\left\{\frac{d}{dt}
	\left(\langle n\rangle^{(i,\text{broad,L})}
	\right)+\frac{\Gamma_{\text{L}}}{\Gamma}\frac{d}{dt}\langle 
	n\rangle^{(i,\text{ren})}
	\right\}. 
\label{eq_second_current}
\end{equation}  
Again, we have written the result in such a form that an identification of
the pumping mechanism is straightforward.
The first term of Eq.~(\ref{eq_second_current}) contains the 
contribution due to the correction of the 
average dot occupation induced by the tunnel coupling to the left lead. 
Intuitively the finite-life-time broadening due to the 
coupling to the left lead is associated with 
tunnel processes of electrons through the left barrier.
Any change in the life-time broadening due to coupling to the left lead will,
therefore, result in a current through the left barrier only. 
As a result, this first term contains a total time derivative, 
and as parameters are periodically changing in time, it will not lead to a
net pumped charge after the full pumping cycle. 
We conclude that changing the life-time broadening of the dot level does not 
contribute to adiabatic pumping.
The second term has the same structure as the zeroth-order contribution,
Eq.~(\ref{eq_second_current}).
It can be understood as the correction term introduced by renormalizing the
position of the dot level, which
 may be time dependent via time-dependent 
tunnel couplings or a time-dependent gate voltage.
Again, the charge transferred in/out of the quantum dot splits into two
currents to or from both leads with relative weight $\Gamma_{\alpha}/\Gamma$.
Now, even if the dot level is constant and only both the tunnel couplings 
$\Gamma_{\alpha}$ are varying in time, a finite
charge can be pumped by means of level renormalization.

It is useful to compare these findings with a perturbation expansion
of the DC current driven by a DC transport voltage.
In lowest (first) order, current is carried by sequential-tunneling processes.
A systematic calculation of the second-order linear conductance~\cite{braggio} 
shows that quantum fluctuation due to tunneling give rise to three different
types of correction terms. 
The first one, which dominates the linear conductance in the Coulomb-blockade
regime away from resonance, is due to cotunneling. 
One way to depict cotunneling is to understand it as transport through the 
finite-life-time broadened dot level.
It would, thus, correspond to the first term of Eq.~(\ref{eq_second_current}).
Close to resonance, however, there are two more corrections to the 
sequential-tunneling linear conductance.
They can be cast as sequential tunneling but with renormalized level
position, as discussed above, or with renormalized tunnel coupling strength.
For the DC current, all these three contributions are present at the same
time, which makes it challenging to identify them separately in an experiment.
For the adiabatically pumped charge, where correction terms associated with a 
renormalization of the tunnel couplings and level-broadening effects vanish, 
the situation is distinctively different.
Studying adiabatic pumping is, therefore, a convenient tool to access the 
energy-level renormalization.
This is most dramatic in the case when the zeroth-order pumped current is zero,
i.e., when pumping is done by changing both tunnel couplings.
In this case, the {\em dominant} contribution to the pumped charge is due to
time-dependent level renormalization.

\subsection{Weak pumping}

When writing the pumped charge, we report as indices, in the following, the 
particular choice of 
pumping parameters it refers to. For example if the pumping 
fields are  $\Gamma_{\text{L}}$ and $\epsilon$, we indicate the charge 
as $Q_{\Gamma_{\text{L}}, \epsilon}$.
We now concentrate on weak pumping.  We write the time-dependent parameters in 
the form $\epsilon(t)=\bar{\epsilon}+\Delta\epsilon(t)$ and 
$\Gamma_{\alpha}(t)=\bar{\Gamma}_{\alpha}+\Delta\Gamma_{\alpha}(t)$ 
(with $\Delta\epsilon(t)$ and $\Delta\Gamma_{\alpha}(t)$ having zero 
time average) and expand 
the current up to bilinear response in the time-dependent part of the 
parameters. 
Choosing  $\epsilon$ and $\Gamma_{\text{L}}$ as pumping 
parameters we obtain up to first order in $\Gamma$:
\begin{equation}\label{eq_charge0_Ufinite}
Q_{\Gamma_{\text{L}},\epsilon} =  -e\frac{\bar{\Gamma}_{\text{R}}}
{\bar{\Gamma}^2}\eta_1  \frac{d}{d\bar{\epsilon}} \left[
\bar{\langle n\rangle}^{(i,0)} 
+\sigma\left(\bar{\epsilon},\bar{\Gamma},U\right)
\frac{d}{d\bar{\epsilon}}\bar{\langle n\rangle}^{(i,0)} \right],
\end{equation}
where the prefactor $\eta_{1}$ characterizes the amplitudes of the pumping 
parameters as well as their relative phase:
\begin{equation}
\nonumber
\eta_{1}=\int_{0}^{\mathcal{T}}
\frac{\partial\Delta\epsilon}{\partial t}\Delta\Gamma_{\text{L}} dt \ .
\end{equation}
It was already pointed out for a noninteracting system in Ref. 
\onlinecite{brouwer} that in the limit of 
weak, adiabatic pumping the pumped charge is proportional to the surface 
enclosed in parameter space during one pumping cycle, which in this case is 
equal to $\eta_{1}$.
Furthermore, $\bar{\langle n\rangle}^{(i,0)}$ is the 
instantaneous occupation of the dot computed with the time-dependent 
parameters taken at their time-average value, and 
$\bar{\Gamma}=\bar{\Gamma}_{\text{L}}+\bar{\Gamma}_{\text{R}}$.
The first term inside the brackets is the zeroth-order-$\Gamma$ contribution 
to the charge and, therefore, it is the dominant one. It has two peaks  
as a function of the average level position, which are located, in the 
limit $U\gg k_{\text{B}} T$, at $\bar{\epsilon}= 
-U-k_{\text{B}} T\ln (2)$ and $\bar{\epsilon}=k_{\text{B}} T\ln (2)$.
The second term is first order in $\Gamma$, stems from level 
renormalization, and  vanishes in the non-interacting case.
The first-order-${\Gamma}$ correction tends to decrease the distance between 
the two resonances.

\begin{figure}
\includegraphics[width=3.in]{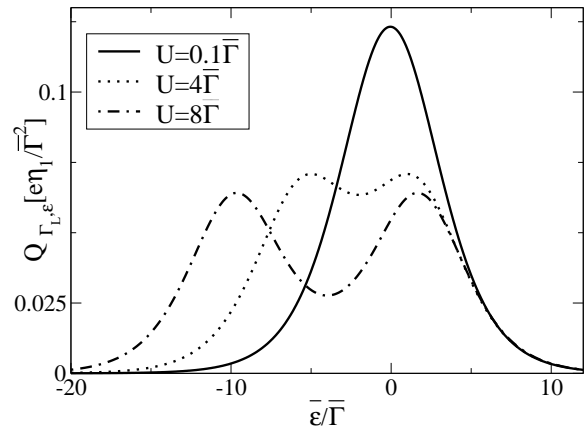}
\caption{
Pumped charge up to first order $\Gamma$ in units of $e\eta_1/\bar{\Gamma}^2$ 
as a function of 
  the time-average level position in units of $\bar{\Gamma}$ for different 
values of $U$. Pumping parameters are $\epsilon$ and $\Gamma_{\text{L}}$. 
The temperature is  $k_{\text{B}} T=2\bar{\Gamma}$.
\label{fig_charge_level}}
\end{figure}
In Fig. \ref{fig_charge_level} the pumped charge of 
Eq.~(\ref{eq_charge0_Ufinite}) is shown in units of $e\eta_1/\bar{\Gamma}^2$ 
as a function of the time-average value of the level position, for different 
strengths of the interaction. 
 The two peaks are directly related to 
transitions between singly-occupied and empty dot and between doubly- and 
singly-occupied dot. 
The shift by $k_{\text{B}} T\ln (2)$ of the peak positions is of combinatoric 
origin and reflects the fact that the probability of single occupation of the 
dot is increased due to the two spin states which lead to the same occupation 
number.
Both peaks contribute with the same sign, as expected since the underlying
pumping mechanism is the same for both resonances: in both cases the dot 
filling increases (decreases) when the level position is decreased (increased).

We now focus our attention on pumping with the two tunnel-coupling strengths
$\Gamma_{\text{L}}$ and $\Gamma_{\text{R}}$. In this case, there is no 
zeroth-order-$\Gamma$ contribution to the pumped charge, as discussed above.
The contribution to first order in $\Gamma$ reads
\begin{equation}
Q_{\Gamma_{\text{L}},\Gamma_{\text{R}}}^{} = e\frac{\eta_2}{\bar{\Gamma}^2}
\frac{d}{d\bar{\epsilon}}\left(\bar{\langle n\rangle}^{(i,0)}\right)
                              \sigma\left(\bar{\epsilon},\bar{\Gamma},U\right)
,
\end{equation}
where $\eta_{2}=\int_{0}^{\mathcal{T}}\frac{\partial\Delta\Gamma_{\text{L}}}
{\partial t}\Delta\Gamma_{\text{R}}dt$ 
accounts for the pumping-parameter
amplitudes and their relative phase
as discussed in detail before for the quantity $\eta_1$ . 
\begin{figure}
\includegraphics[width=3.15 in]{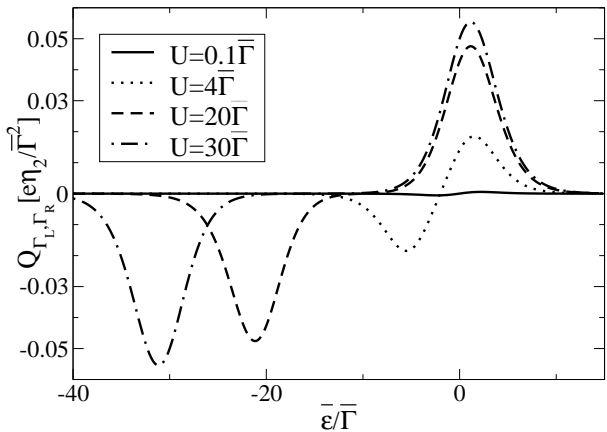}
\caption{Pumped charge up to  first order $\Gamma$ in units of 
$e\eta_2/\bar{\Gamma}^2$ as a function of 
  the time-average level position in units of $\bar{\Gamma}$ for different 
values of $U$. Pumping parameters are $\Gamma_{\text{L}}$ and 
$\Gamma_{\text{R}}$. The temperature is  $k_{\text{B}} T=2\bar{\Gamma}$.
\label{fig_secondorder}}
\end{figure}
The result for the pumped charge as a function of the level position is shown 
in Fig. \ref{fig_secondorder}. The solid line shows the result for very small 
interaction. As expected, it tends to zero, because the level renormalization 
vanishes. In the presence of interaction two peaks appear, which separate for 
increasing $U$. 
The two peaks are related to the two resonances at the level positions. 
They contribute with opposite sign.
This reflects the opposite sign of the level renormalization for the 
two resonances.
We remind that the first non-vanishing contribution of the 
perturbation expansion to the charge pumped through the dot by periodic change 
of the tunnel barriers is uniquely due to the effect of level renormalization.
The height of the peaks increases for increasing $U$, growing 
logarithmically for large $U$.
Eventually, this increase will be cut-off by the bandwidths $D$, which we 
here chose to be infinite.
The different sign of the pumped charge for the two resonances could serve as a
signature to distinguish level-renormalization-induced pumping from parasitic 
peristaltic pumping due to cross capacitances of the gates modulating the 
tunnel couplings to the quantum dot.

\section{Conclusions}
\label{conclusions}

We have presented a perturbative approach in tunneling to 
adiabatic pumping through interacting quantum dots. In particular, a 
general diagrammatic technique to perform the adiabatic expansion has been 
developed. This technique has been applied to compute the pumped charge 
through a single-level quantum dot at temperatures much higher than the Kondo 
temperature. Two pumping schemes have been considered: pumping with the level 
position and one tunnel barrier, and pumping with the two barriers. When 
pumping with the level position and one tunnel barrier, 
the dominant mechanism of the adiabatic pump works analogously to a 
peristaltic pump. The next-order correction is related to the level 
renormalization induced by the interplay of Coulomb repulsion and electron 
tunneling. 
The situation is far more interesting for the case of pumping with the two 
barriers. With this pumping scheme there is no pumping in lowest order in the 
tunnel coupling, and the first 
non-vanishing contribution is due to the time-dependent level renormalization.
Hence, we have demonstrated the importance of level-renormalization effects in 
pumping 
through interacting quantum dots. In particular, our 
results suggest that adiabatic pumping can be used to gain experimental access 
to the level renormalization in quantum dots.

We acknowledge financial support by the European Community via grants RTNNANO 
and MIUR-PRIN, the DFG via SFB491, and the NSF under grant No.~PHY99-0794.

\begin{appendix}

\section{Examples of diagrams}\label{ap_example}

\begin{figure}[b]
\includegraphics[width=3.in]{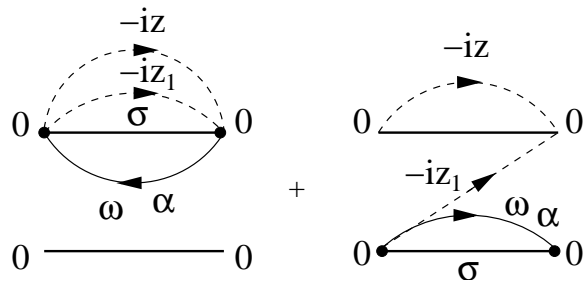}
\caption{Diagrams contributing to the first-order-$\Gamma$ part of the 
adiabatic correction to the matrix element $\left(\mathbf{W}_{t}\right)_{0,0}$.
All appearing reservoir and spin indices $\alpha$, $\sigma$ are to be summed 
over and the energy $\omega$ is to be integrated over.
\label{fig_ad_example}}
\end{figure}

In this section we show how to apply the diagrammatic rules for the 
matrix element $(\mathbf{W}_{t})_{0,0}$.
\\
We start with the instantaneous term to lowest order in the tunnel
coupling, $(\mathbf{W}_{t}^{(i,1)}(z))_{0,0}$.
The corresponding diagrams are shown in Fig. ~\ref{fig_ad_example}.
Two topologically different diagrams contribute, and each of them has to be 
summed over the spin index $\sigma$ and the lead index $\alpha$, and to be
integrated over $\omega$. 
We obtain:
\begin{equation}
\left(\mathbf{W}_{t}^{(i,1)}(z)\right)_{0,0}= -i
\sum_{\sigma,\alpha}\int\frac{d\omega}{2\pi}\left[ \frac{\Gamma_\alpha 
f\left(\omega\right)}{\omega-\epsilon+iz}
+ \frac{\Gamma_\alpha 
f\left(\omega\right)}{\epsilon-\omega+iz} \right] \, .
\nonumber
\end{equation}
By letting $z=0_+$ and making use of $1/(x+i0_+) = P/x - i \pi \delta(x)$, 
where $P$ indicates
Cauchy's principal value, we get
\begin{eqnarray}
\left(\mathbf{W}_{t}^{(i,1)}\right)_{0,0} &=& - 2\Gamma f\left(\epsilon \right)
\, ,
\nonumber
\\
\left(\partial \mathbf{W}_{t}^{(i,1)} \right)_{0,0} &=& -\frac{2\Gamma}{\pi} 
\frac{d}{d\epsilon} \int_{\mathrm{P}} d\omega 
\frac{f\left(\omega\right)}{\omega-\epsilon}
\, .
\nonumber
\end{eqnarray}

For the adiabatic correction we need to introduce additional external 
frequency lines according to the rules 7a and 8a 
(see Fig.\ref{fig_ad_example}). The additional line of
rule 8a, going from the left corner of the diagram with no vertex to the right
corner, does not contribute in this case and we have omitted drawing it. 
The evaluation of these diagrams leads us to the result
\begin{eqnarray*}
\left(\mathbf{W}_{t}^{(a,1)}\right)_{0,0} & = &-\frac{\frac{d\Gamma}{dt}}{\pi} 
\frac{d}{d\epsilon} \int_{\mathrm{P}} d\omega 
\frac{f\left(\omega\right)}{\omega-\epsilon}
- \frac{\Gamma \frac{d\epsilon}{dt}}{\pi} \frac{d^2}{d\epsilon^2} 
\int_{\mathrm{P}} d\omega \frac{f\left(\omega\right)}{\omega-\epsilon}
\end{eqnarray*}

\begin{figure*}
\includegraphics[width=7.in]{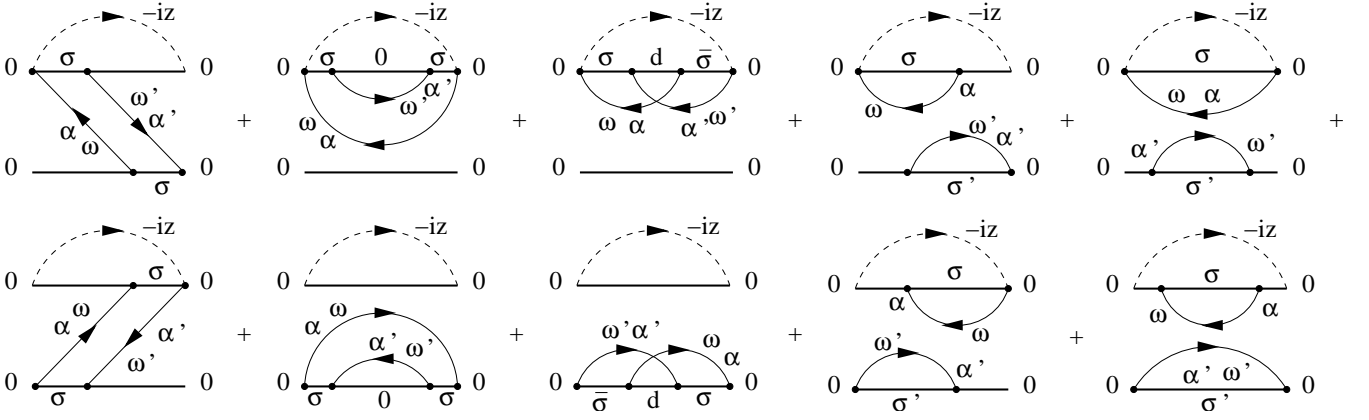}
\caption{Diagrams contributing to the second-order-$\Gamma$ instantaneous 
part of the matrix element $\left(\mathbf{W}_{t}\right)_{0,0}$. All appearing 
reservoir and spin indices $\alpha$, $\alpha'$, $\sigma$, $\sigma'$ are to be 
summed over and the energy $\omega$ is to be integrated over.
\label{fig_inst_example2}}
\end{figure*}

 We now calculate the second-order-$\Gamma$ contribution to 
the same matrix element of the instantaneous kernel.
All diagrams contributing to the matrix element 
$(\mathbf{W}_{t}^{(i,2)})_{0,0}$ are depicted in 
Fig.~\ref{fig_inst_example2}. We sum over all appearing indices 
$\alpha, \alpha', \sigma, \sigma'$. The variables $\bar{\sigma}$ denotes the
opposite spin of $\sigma$.
As an example, we report the result for the sum of the first diagram of the
first and the second line (after setting $z=0_+$):
\begin{eqnarray*}
\frac{\Gamma^2}{\pi}\left[2f(\epsilon) \frac{d}{d\epsilon}\int_{\mathrm{P}}
d\omega\frac{f(\omega)}{\omega-\epsilon}-\frac{d}{d\epsilon}
\int_{\mathrm{P}}d\omega\frac{\left(f(\omega)\right)^2}{\omega-\epsilon}\right]
\, . 
\end{eqnarray*}
To obtain the full second-order contribution 
$(\mathbf{W}_{t}^{(i,2)})_{0,0}$ we need to evaluate the remaining 
diagrams in Fig.~\ref{fig_inst_example2} along the same lines as discussed 
in this Appendix.

\section{Occupation probabilities}
\label{ap_results}

Some intermediate results were not presented in the main part of this 
article as they were lengthy or 
not immediately necessary for the interpretation of the pumped current. 
Here we discuss in detail the first corrections to the occupation 
probabilities. The corrections to the occupation probabilities are used for 
the evaluation of 
Eq.~(\ref{expcur})  but do not appear directly in the results for the pumped 
current. We find for $\mathbf{p}_{t}^{(a,-1)}$, the first order adiabatic 
correction in
 minus first order in $\Gamma$:
\begin{equation}
\mathbf{p}_{t}^{(a,-1)}=-\frac{d\mathbf{p}_{t}^{(i,0)}}{dt}
\frac{1}{2\Gamma}\frac{1}{\left[1+f\left(\epsilon\right)-
f\left(\epsilon+U\right)\right]}.\nonumber
\end{equation}
The adiabatic correction of the occupation probability is 
proportional to the time-derivative of $\mathbf{p}_{t}^{(i,0)}$, and is 
therefore an eigenvector of the matrix $\mathbf{W}_{t}^{(i,1)}$. 
The sign of this correction depends on the sign of the time derivative of 
$\epsilon\left(t\right)$.

The first-order-$\Gamma$ correction to the instantaneous occupation 
probability is:
\begin{eqnarray*}
\mathbf{p}_{t}^{(i,1)} & = & \frac{d\mathbf{p}_{t}^{(i,0)}}{d\epsilon}
\sigma\left(\epsilon,\Gamma,U\right)\\
 & + & \left(2-\langle n\rangle^{(i,0)}\right)
\phi'\left(\epsilon\right)\left(-1,\frac{1}{2},\frac{1}{2},0\right)
^{\mathrm{T}}
\\ & + &\langle n\rangle^{(i,0)}
\phi'\left(\epsilon+U\right)\left(0,-\frac{1}{2},-\frac{1}{2},1\right)
^{\mathrm{T}}\,.
\end{eqnarray*} 
It consists of a part due to level renormalization (first row) and a part due 
to level broadening (second and third row). The correction due to level 
renormalization affects all four probabilities in the same functional way.\\ 
The correction due to broadening has two contributions. The first one is 
related to the broadening due to fluctuations between empty and 
singly-occupied dot.
It is zero in the 
case that the dot is doubly occupied and largest when the dot is empty. The  
second contribution is related to the broadening due to fluctuations between
singly- and doubly-occupied dot.
It is zero when the dot is empty and largest when the dot is doubly occupied.

\end{appendix}


\begin{thebibliography}{24}

\bibitem{geometric} B. L. Altshuler and L. I. Glazman, Science {\bf 283}, 1864 
                 (1999); Y. Makhlin and A. D. Mirlin, Phys. Rev. Lett. 
                 {\bf 87}, 276803 (2001); H.-Q. Zhou, S. Y. Cho, and 
                 R. H. McKenzie, Phys. Rev. Lett. {\bf 91}, 186803 (2003).

\bibitem{thouless}  D. J. Thouless, Phys. Rev. B \textbf{27}, 6083 (1983). 


\bibitem{pothier} H. Pothier, P. Lafarge, C. Urbina, D. Esteve, and M.H. 
             Devoret, Europhys. Lett. {\bf 17}, 249 (1992).


\bibitem{brouwer} P. W. Brouwer, Phys. Rev B {\bf 58}, R10135 (1998).


\bibitem{zhou} F.~Zhou, B.~Spivak, and B.~Altshuler, Phys. Rev. Lett.
        {\bf 82}, 608 (1999).


\bibitem{woelfle} Y. Levinson, O. Entin-Wohlman, and P. W\"olfle, 
Phys. Rev. Lett. {\bf 85}, 634 (2000);
O. Entin-Wohlman, Y. Levinson, and P. W\"olfle, Phys. Rev. B {\bf 64}, 195308 
(2001).

\bibitem{levitov} L.~S.~Levitov, cond-mat/0103617 (2001).

\bibitem{levinson} O.~Entin-Wohlman, A.~Aharony, and
        Y. Levinson, Phys. Rev. B {\bf 65}, 195411 (2002).

\bibitem{aleiner1}
        I. L. Aleiner, B. L. Altshuler, and A. Kamenev,
        Phys. Rev. B {\bf 62}, 10373 (2000).
\bibitem{moskalets} M.~Moskalets and M.~B\"uttiker, Phys. Rev. B {\bf 66},
035306 (2002).
\bibitem{butti} M.~Moskalets and M.~B\"uttiker, Phys. Rev. B {\bf 64},
201305 (2001).



\bibitem{mucciolo}
        E. R. Mucciolo, C. Chamon, and  C. M. Marcus, Phys. Rev. Lett.
        {\bf 89}, 146802 (2002).

\bibitem{spinpump}
        M. Governale, F. Taddei, and R. Fazio, Phys. Rev. B {\bf 68},
        155324 (2003).

\bibitem{quantumhall}
M. Blaauboer, Phys. Rev. B {\bf 68}, 205316 (2003).

\bibitem{wangapl}
J. Wang, Y. Wei, B. Wang, and H. Guo, Appl. Phys. Lett. {\bf 79}, 3977 (2001).

\bibitem{blaauboer} M. Blaauboer, Phys. Rev. B {\bf 65}, 235318 (2002).

\bibitem{andreevinterf}  F. Taddei, M. Governale, and R. Fazio, Phys. Rev. B 
\textbf{70}, 052510 (2004).


\bibitem{shutenko}  T. A. Shutenko, I. L. Aleiner, and B. L. Altshuler, 
                 Phys. Rev. B {\bf 61}, 10366 (2000).

\bibitem{vavilov}  M. G. Vavilov, V. Ambegaokar, and I. L. Aleiner, 
                  Phys Rev. B {\bf 63}, 195313 (2001).

\bibitem{polianski}  M. L. Polianski and P. W. Brouwer, 
                 J. Phys. A: Math. Gen. {\bf 36}, 3215 (2003).



\bibitem{switkes} M. Switkes, C. M. Marcus, K. Campman, and A. C. Gossard, 
                    Science {\bf 283}, 1905 (1999).

\bibitem{dicarlo} L. DiCarlo, C. M. Marcus, and J. S. Harris, Jr., 
              Phys. Rev. Lett. {\bf 91}, 246804 (2003).

\bibitem{watson} S. K. Watson, R. M. Potok, C. M. Marcus, and V. Umansky, 
              Phys. Rev. Lett {\bf 91}, 258301 (2003).


\bibitem{ahlers} N. E. Fletcher, J. Ebbecke, T. J. B. M. Janssen, F. J. Ahlers,
           M. Pepper, H. E. Beere, and D. A. Ritchie, Phys. Rev. B {\bf 68}, 
           245310 (2003); J. Ebbecke, N. E. Fletcher, T. J. B. M. Janssen, 
           F. J. Ahlers, M. Pepper, H. E. Beere, and D. A. Ritchie, Appl. Phys.
           Lett. {\bf 84}, 4319 (2004).



\bibitem{aleiner} I. L. Aleiner and A. V. Andreev, Phys. Rev. Lett. 
                      \textbf{81}, 1286 (1998).

\bibitem{brouwer2} P. W. Brouwer, A. Lamacraft, and K. Flensberg, Phys. Rev. B 
                     {\bf 72}, 075316 (2005).

\bibitem{aono}       T. Aono, Phys. Rev. Lett. \textbf{93}, 116601 (2004).


\bibitem{cota}  E. Cota, R. Aguado, and G. Platero, Phys. Rev. Lett. {\bf 94}, 
                107202 (2005); E. Cota, R. Aguado, and G. Platero, Phys. Rev. 
                Lett. {\bf 94}, 229901(E) (2005).
\bibitem{citro}  R. Citro, N. Andrei, and Q. Niu, Phys. Rev. B {\bf 68}, 
                     165312 (2003).

\bibitem{green} J. Splettstoesser, M. Governale, J. K\"onig, and R. Fazio, 
               Phys. Rev. Lett. \textbf{95}, 246803 (2005).


\bibitem{sela}  E. Sela and Y. Oreg, Phys. Rev. Lett. {\bf 96}, 166802 (2006).

\bibitem{buttiker} 
        M. B\"uttiker, H. Thomas, and A. Pr\^etre, Z. Phys. B: 
        Condens. Matter \textbf{94}, 133 (1994).

\bibitem{koenig} J. K\"onig, H. Schoeller, and G. Sch\"on, Phys. Rev. Lett. 
  {\bf76}, 1715 (1996); 
  J. K\"onig, J. Schmid, H. Schoeller, and G. Sch\"on, Phys. Rev. B {\bf 54}, 
  16820 (1996);
  H. Schoeller, in \textit{Mesoscopic Electron Transport}, edited by L.L. Sohn,
  L.P. Kouwenhoven, and  G. Sch\"on (Kluwer, Dodrecht, 1997);
  J. K\"onig, \textit{Quantum Fluctuations in the Single-Electron Transistor}
  (Shaker, Aachen, 1999).


\bibitem{comment} 

This can be easily proven within a diagrammatic real-time approach, that 
has been used to study a variety of quantum effects of transport \cite{koenig},
 and that will be employed later in this paper.



 
\bibitem{poorman} F. D. M. Haldane, Phys. Rev. Lett. {\bf 40}, 416 (1978); 
                     A. C. Hewson, \textit{The Kondo Problem to Heavy Fermions}
                     (Cambridge University Press, Cambridge, England, 1993).

\bibitem{braggio} A. Braggio, J. K\"{o}nig, and R. Fazio, Phys. Rev. Lett.
{\bf 96}, 026805 (2006).


\end{thebibliography}
\end{document}